\documentclass[10pt,twocolumn,aps,pra,showpacs,superscriptaddress,floatfix,longbibliography,nofootinbib]{revtex4-2}

\usepackage{graphicx}
\usepackage{subfigure}
\usepackage[T1]{fontenc}
\usepackage{graphicx}
\usepackage[utf8]{inputenc}
\usepackage{amsmath, amsthm, amssymb,amsfonts}
\usepackage{color}
\usepackage{psfrag}
\usepackage{epsfig}
\usepackage{bbm}
\usepackage{bm}
\usepackage[colorlinks=true,citecolor=blue,urlcolor=magenta]{hyperref}
\usepackage[normalem]{ulem}
\usepackage{epstopdf}
\usepackage{graphicx}
\usepackage{stackengine,xcolor}
\usepackage{comment}

\definecolor{nred}{rgb}{0.9,0.1,0.1}
\definecolor{nblack}{rgb}{0,0,0}
\definecolor{nblue}{rgb}{0.2,0.2,0.8}
\definecolor{ngreen}{rgb}{0.2,0.6,0.2}

\newcommand{\blu}{\color{nblue}}

\usepackage{etoolbox}
\usepackage{bbold} 

\newcommand{\beq}{\begin{eqnarray}}
\newcommand{\eeq}{\end{eqnarray}}

\newcommand{\id}{\openone}

\newcommand{\rab}{{\varrho^{\text{AB}}}}

\DeclareMathOperator{\tr}{tr}

\theoremstyle{definition}

\begin{document}

\title{Investigating an approach of robustly self-testing two-qubit entangled states}

\author{Chan-Ching Lien} 
\affiliation{Department of Physics, National Cheng Kung University, Tainan 701, Taiwan}
\affiliation{Center for Quantum Frontiers of Research \& Technology (QFort), National Cheng Kung University, Tainan 701, Taiwan}
\affiliation{Department of Physics, National Chung Hsing University, Taichung 402, Taiwan}

\author{Shin-Liang Chen}
\email{shin.liang.chen@email.nchu.edu.tw}
\affiliation{Department of Physics, National Chung Hsing University, Taichung 402, Taiwan}
\affiliation{Physics Division, National Center for Theoretical Sciences, Taipei 106319, Taiwan}
\affiliation{Center for Quantum Frontiers of Research \& Technology (QFort), National Cheng Kung University, Tainan 701, Taiwan}

\date{ \today}

\begin{abstract}
In a recent paper [Quantum \textbf{5}, 552 (2021)], the authors proposed a framework for robustly self-testing steerable quantum assemblages. In this work, we apply their method to the scenario of self-testing two-qubit entangled quantum states. The new bounds on the fidelity with the reference states are compared with other methods.
\end{abstract}

\maketitle

\section{Introduction}
Since Bell's seminal paper~\cite{Bell64}, quantum nonlocality~\cite{Brunner14} plays an essential role in the field of quantum information science. Fundamentally, nonlocality changed our understanding of nature. In practice, it provides a powerful tool called \emph{device-independent quantum certification}~\cite{Acin07,Scarani12}. The device-independent protocol certifies quantum properties without making assumptions on the underlying states nor the involving measurement devices. Such a certification includes entanglement certification~\cite{Wiseman07,Moroder13}, steering certification~\cite{Wiseman07,Cavalcanti16,CBLC16}, incompatibility certification~\cite{Wolf09,CMBC2021},  dimension witnesses~\cite{Gallego10}, etc.

The most extreme case in device-independent (DI) quantum certification is identifying the underlying states and measurements themselves. The so-called \emph{self-testing} protocol was firstly introduced by Mayers and Yao~\cite{Mayers98,Mayers04}. Since then, many research results on self-testing have been proposed, ranging from self-testing entangled state~\cite{McKague11,Yang14,Kaniewski16,Coladangelo17}, incompatible measurements~\cite{Yang14,Kaniewski2017,Bowles18PRA}, steerable assemblages~\cite{Chen2021robustselftestingof}, preparation and measurements~\cite{Tavakoli18b,Miklin2021universalscheme,Chen2024semi},  entangled measurements~\cite{Renou18}, and so on
(see Ref.~\cite{Supic19} for a comprehensive review of self-testing).

Self-testing has been studied mainly through the following approaches: The typical analytical approach~\cite{Mayers98,Mayers04,McKague11}, the swap method~\cite{Yang14,Bancal15}, and the operator inequality method~\cite{Kaniewski16}. Different approaches have their features and benefits. The typical analytical approach is universal in general and can be generalized from simple system to multipartite system (e.g.,  Ref.~\cite{Bowles18}). The computation of the swap method is carried out by semidefinite programming and often provides more robust results than the analytical approach (e.g.,  Ref.~\cite{Yang14,Bancal15}). The operator inequality method is analytical, but the concept is totally different from the standard analytical approach. The operator inequality method often provides the best and remarkably robust results than the other two methods (e.g., Ref.~\cite{Kaniewski16}).

\begin{figure*}
\begin{minipage}[c]{.49\textwidth}
\includegraphics[width=8.8cm]{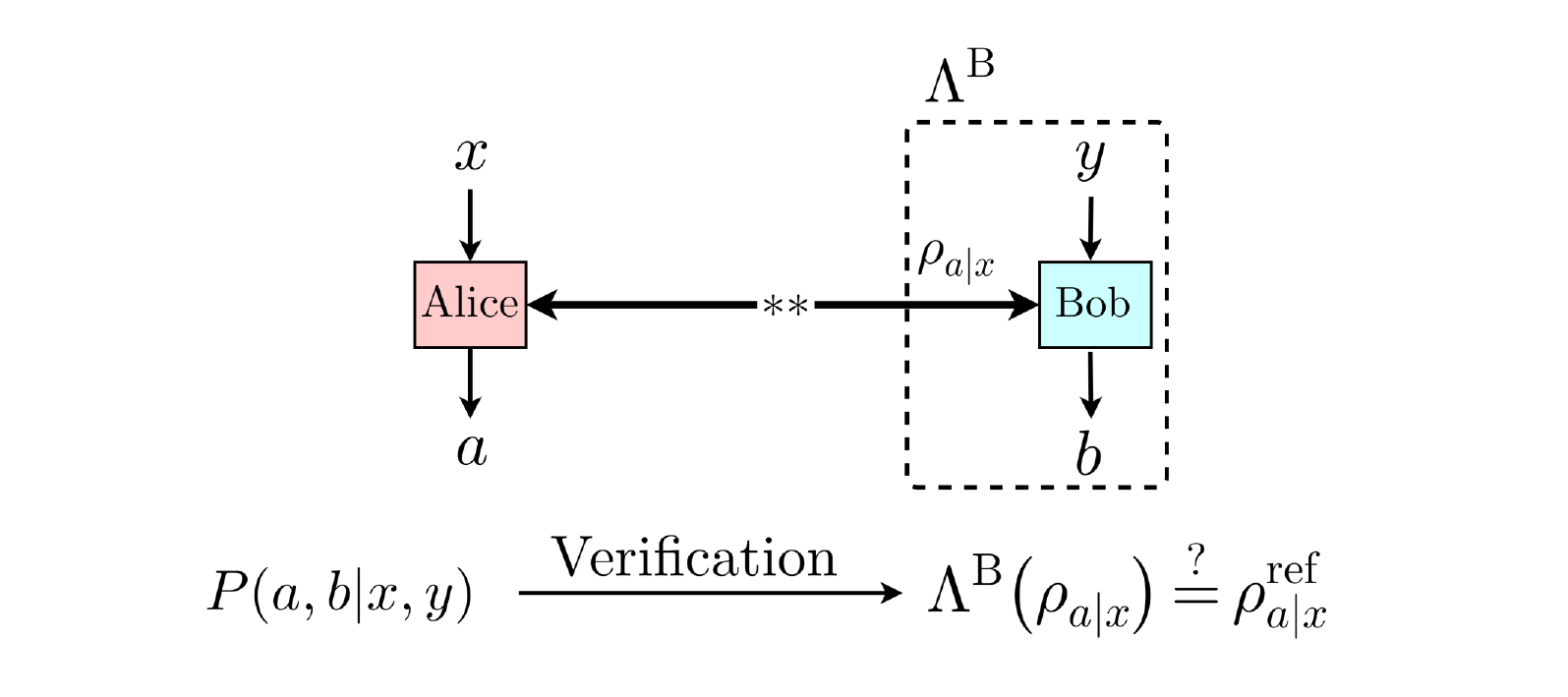}\\
\centering
\text{(a)}
\end{minipage}
\begin{minipage}[c]{.49\textwidth}
\includegraphics[width=8.8cm]{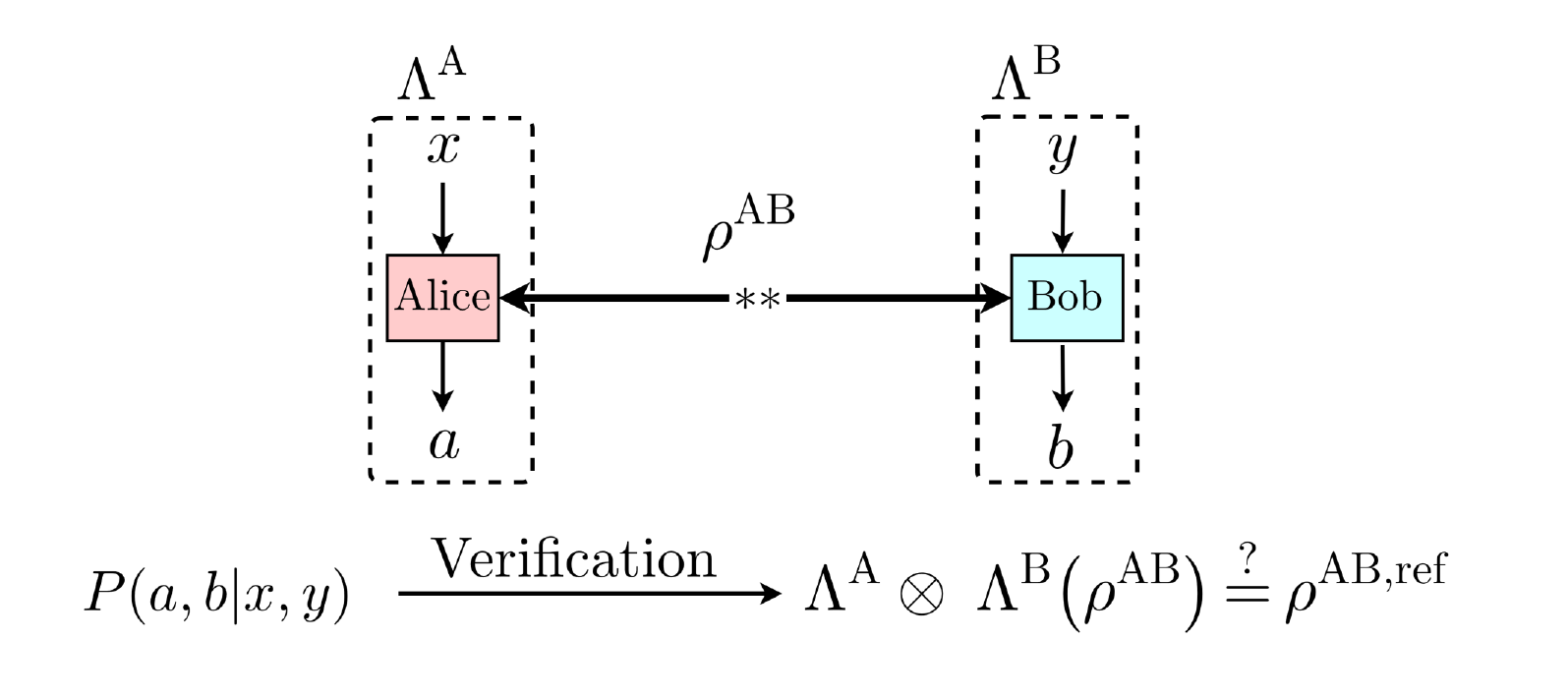}\\
\centering
\text{(b)}
\end{minipage}
\caption{
Overview of this work. Given observed statistics $P(a,b|x,y)$ in a Bell scenario (see Sec. \ref{Sec_Bell}), (a) the authors of \cite{Chen2021robustselftestingof} proposed a method to certify if the underlying steerable assemblage $\rho_{a|x}$ (up to a local completely positive and trace-preserving (CPTP) map $\Lambda^{\rm B}$) is identical to the reference assemblage $\rho_{a|x}^{\rm ref}$ , while here, (b) we apply their method to verify if the underlying shared state $\rho^{\rm AB}$ (up to a local CPTP map $\Lambda^{\rm A}\otimes\Lambda^{\rm B}$) is identical to the reference state $\rho^{\rm AB, ref}$. 
}
\label{Fig_overview}
\end{figure*}

Recently, a new approach was introduced to self-test steerable quantum assemblages~\cite{Chen2021robustselftestingof}. Briefly speaking, the authors of Ref.~\cite{Chen2021robustselftestingof} follow the idea of the swap method, but they construct the local isometry from the viewpoint of channel-state duality. In this work, we apply the method to the scenario of self-testing two-qubit entangled quantum states and compare the results with other approaches. The overview picture is shown in Fig.~\ref{Fig_overview}.

The article is organized as follows. First, we briefly review the Bell scenario. Then, in Section~\ref{Sec_ST_states}, we give our general framework for self-testing of entangled states. After that, we consider a typical example of self-testing the two-qubit maximally entangled state in Section~\ref{Sec_ST_MES}. We also consider a more general example of self-testing a family of pure entangled states in Sec.~\ref{Sec_ST_PES}. Finally, we address some open problems and future outlook in Sec.~\ref{Sec_summary}.

\section{The Bell scenario}\label{Sec_Bell}

The Bell scenario considered in this work is composed as follows. A pair of particles are respectively sent to two parties, called Alice and Bob, who perform measurements on the received particle. Alice's measurement choices and outcomes are labeled respectively by $x$ and $a$ while Bob's are labeled by $y$ and $b$. After repeating the experiment for many rounds, they can obtain probabilities of measurement outcomes conditional on the measurement choices, denoted as $P(a,b|x,y)$. The collection of $P(a,b|x,y)$ is called the \emph{correlation}, denoted as $\mathbf{P}$, i.e., $\mathbf{P}:=\{P(a,b|x,y)\}$.

If Alice and Bob live in a world described by local causality, the correlation admits a local-hidden-variable model~\cite{Bell64,Brunner14}. That is,
\begin{equation}
P(a,b|x,y) = \sum_\lambda P(\lambda)P(a|x,\lambda)P(b|y,\lambda)
\label{Eq_local_correlation}
\end{equation}
for all $a,b,x,y$, where $\lambda$ are hidden variables or the so-called \emph{shared randomness}. In this case, we say that the correlation between Alice and Bob is \emph{local}. A \emph{Bell inequality} is a linear combination of $P(a,b|x,y)$ and it is a way to demonstrate the limitation that all local correlations can achieve. In general, it is expressed as
\begin{equation}
\vec{\beta}\cdot \mathbf{P}:=\sum_{a,b,x,y} \beta_{a,b}^{x,y}P(a,b|x,y)\leq L,
\label{Eq_BI}
\end{equation}
where $\beta_{a,b}^{x,y}$ are some real numbers, the set of which defines a Bell inequality. $L$ is called the \emph{local bound} for the Bell inequality. 

In quantum world, Alice and Bob's pair is described by a quantum state $\rab$. The measurements performed by them are described by positive operator-valued measures (POVMs) $E_{a|x}$ and $E_{b|y}$. The correlation now is obtained by the Born rule:
\begin{equation}
P(a,b|x,y) = \tr(E_{a|x}\otimes E_{b|y}\rab)
\label{Eq_pabxy}
\end{equation} 
for all $a,b,x,y$. The correlation obtained in quantum world is called \emph{quantum correlation}, and we can see that it is different from the local correlation in Eq.~\eqref{Eq_local_correlation}. In fact, such a difference can cause the violation of a Bell inequality \eqref{Eq_BI}. That is, there exists a quantum state $\rab$ and measurements $E_{a|x}, E_{b|y}$ such that Eq.~\eqref{Eq_BI} does not hold. A correlation $\mathbf{P}$ that yields the violation of a Bell inequality is referred to as a \emph{nonlocal correlation}. Similar to the local case, quantum correlations also have limitations on achieving a Bell value. That is,
\begin{equation}
\vec{\beta}\cdot \mathbf{P}:=\sum_{a,b,x,y} \beta_{a,b}^{x,y}P(a,b|x,y)\leq Q.
\end{equation}
In this paper, we are interested in the situation of $Q>L$. In self-testing scheme, one tries to identify the quantum systems (states, measurements, assemblages, etc.) themselves through the maximal quantum violation of a Bell inequality, i.e., when observing $\vec{\beta}\cdot \mathbf{P}_{\rm obs}=Q$~\cite{FN1}. We will review the knowledge of self-testing in the coming section.

\section{Self-testing of entangled states: Definition and Formulation}\label{Sec_ST_states}

After reviewing the concept of nonlocality, we can see that the correlation $\{P(a,b|x,y)\}$ and the state $\rab$ can be connected via Eq.~\eqref{Eq_pabxy}.

Namely,  the correlation is obtained by performing Alice's and Bob's measurements on the shared state. The self-testing problem asks the reverse question: When observing a correlation $\{P(a,b|x,y)\}$, can we uniquely identify the underlying state? Since the correlation is the only known information and we do not assume the form nor the dimension of the underlying state, we cannot distinguish a state $\rab$ and its isometry equivalence $(V\otimes W) \rab (V^\dag\otimes W^\dag)$, i.e.,
\begin{equation}
\begin{aligned}
&\tr(E_{a|x}\otimes E_{b|y} \rab) =\\
& \tr\Big[ (VE_{a|x}V^\dag)\otimes(WE_{b|y}W^\dag)(V\otimes W) \rab (V^\dag\otimes W^\dag)\Big].
\end{aligned}
\end{equation}
Therefore, what we actually certify is an equivalence-class state. Defining a \emph{reference} (or \emph{representative}) state $\rho^{\rm ref}$ and associated optimal measurements $E_{a|x}^{\rm opt}, E_{b|y}^{\rm opt}$ yielding to the same correlation, we have
\begin{equation}
\begin{aligned}
\tr &\Big[ (VE_{a|x}V^\dag)\otimes(WE_{b|y}W^\dag) (V\otimes W) \rab (V^\dag\otimes W^\dag)\Big]\\
&= \tr(E_{a|x}^{\rm opt}\otimes E_{b|y}^{\rm opt} \rho^{\rm ref})
\end{aligned}
\end{equation}
Here, $(V\otimes W) \rab (V^\dag\otimes W^\dag) = \rho^{\rm ref}$, or equivalently, by the Stinespring dilation, $\Lambda(\rho) = \rho^{\rm ref}$, where $\Lambda$ is a completely positive and trace-preserving (CPTP) map. Having all the ingredients above, we can define the self-testing of quantum state\cite{Kaniewski16,Supic19}:

\textbf{Definition (self-testing of entangled states)} We say that the observed correlation $P_{\rm obs}(a,b|x,y)$ self-tests the reference state $\rho^{\rm ref}$ if for each state $\rho$ compatible with $P_{\rm obs}(a,b|x,y)$ there exist CPTP maps $\Lambda^{\rm A}$, $\Lambda^{\rm B}$ such that
\begin{equation}
\Lambda^{\rm A} \otimes \Lambda^{\rm B} (\rho)=\rho^{\rm ref}.
\end{equation}
From the resource theory of entanglement, the CPTP map has to be in the form of separable operation $\Lambda^{\rm A}\otimes\Lambda^{\rm B}$ since it cannot increase the degree of entanglement. Otherwise, a global CPTP map can always map the underlying state to the reference one. 

In a practical situation, the underlying system is always suffered from environmental noise, which makes the observed correlation depart from the ideal value. Therefore, we are not able to draw a conclusion on the perfect self-testing. Nevertheless, we can still estimate how close the underlying state and the reference one are. Such a scheme is called \emph{robust self-testing states}~\cite{McKague12_0}, which we define as follows:

\textbf{Definition (robust self-testing of entangled states)} We say that the observed correlation $P_{\rm obs}(a,b|x,y)$ \emph{robustly} self-tests the reference state $\rho^{\rm ref}$ with the fidelity $f$ if for each state $\rho$ compatible with $P_{\rm obs}(a,b|x,y)$ there exist CPTP maps $\Lambda^{\rm A}$, $\Lambda^{\rm B}$ such that
\begin{equation}
F(\Lambda^{\rm A} \otimes \Lambda^{\rm B} (\rho),\rho^{\rm ref}) \geq f,
\label{Eq_Def_ST_state}
\end{equation}
where $F$ is the fidelity between two quantum states. For the reference state being pure (which is the case {\blu in the conventional self-testing scheme~\cite{Supic19}}), we have
\begin{equation}
F = \langle \psi^{\rm ref} | \Lambda^{\rm A} \otimes \Lambda^{\rm B} (\rho) |\psi^{\rm ref} \rangle
\end{equation}
Note that one can treat the above definition as a special case of the definition proposed in Ref.~\cite{Kaniewski16}, where the author also considers the optimization over the CPTP maps $ \Lambda^{\rm A}$ and $ \Lambda^{\rm B}$. Here, in contrast, we only consider the existence of CPTP maps.

Now, we use the channel-state duality to represent the CPTP maps as Choi matrix~\cite{Jamiokowski74,Choi75}. Namely, 
\begin{equation}
\begin{aligned}
&\Lambda^{\rm A} \otimes \Lambda^{\rm B} (\rho^{\rm AB}) =\\
& \tr_{\rm AB}\Big[ \Big( \Omega^{\rm AA'}\otimes \Omega^{\rm BB'}\Big)\Big( (\rho^{\rm AB})^{\mathsf{T}}\otimes\mathbb{1}^{\rm A'B'} \Big)\Big],
\end{aligned}
\label{Eq_duality}
\end{equation}
where $\Omega^{\rm AA'}$ and $\Omega^{\rm BB'}$ are Choi matrices for the CPTP maps $\Lambda^{\rm A}$ and $\Lambda^{\rm B}$, respectively. The action of transpose with respect to the computational basis is denoted as $\mathsf{T}$. The Choi matrix $\Omega^{\rm AA'}$ for $\Lambda^{\rm A}$ is defined as $\Omega:=(\mathsf{id}\otimes \Lambda)|\phi^+\rangle\langle\phi^+|$ with $|\phi^+\rangle:=\sum_i |i\rangle_{\rm A}\otimes |i\rangle_{\rm A}$ being the subnormalized maximally entangled state and $\mathsf{id}$ being the identity operation. A similar definition holds for the other Choi matrix $\Omega^{\rm BB'}$. Because of the relation Eq.~\eqref{Eq_duality}, the problem of searching for CPTP maps $\Lambda^{\rm A}, \Lambda^{\rm B}$ is equivalent to the problem of searching for positive semidefinite matrices $\Omega^{\rm AA'}, \Omega^{\rm BB'}$.

In the following sections, we show that the method of self-testing assemblages in Ref.~\cite{Chen2021robustselftestingof} can be applied to the self-testing of entangled states. That is, we will show how to compute bounds on $f$ in Eq.~\eqref{Eq_Def_ST_state} using the method of Ref.~\cite{Chen2021robustselftestingof}.

\section{Self-testing of the two-qubit maximally entangled state}\label{Sec_ST_MES}
To make readers understand how our method works, we firstly consider the simplest self-testing scenario, i.e., self-testing the two-qubit maximally entangled state.
In this scenario, if Alice and Bob perform measurements in the following bases
\begin{equation}
\begin{aligned}
&A_1 =Z, \quad B_1 =\frac{Z+X}{\sqrt{2}},\\
&A_2=X, \quad B_2=\frac{Z-X}{\sqrt{2}}
\end{aligned}
\end{equation}
on the maximally entangled state
\begin{equation}
|\psi\rangle = \frac{1}{\sqrt{2}}(|00\rangle + |11\rangle,
\end{equation}
then such a strategy can be used for violating the Clauser-Horne-Shimony-Holt (CHSH) inequality~\cite{Clauser69}:
\begin{equation}
I_{\rm CHSH}:= \langle A_1 B_1\rangle + \langle A_1 B_2\rangle + \langle A_2 B_1\rangle - \langle A_2 B_2\rangle\leq 2
\end{equation}
and achieving the maximal quantum violation of $2\sqrt{2}$.

The self-testing issue asks the reverse: Can we identify that the underlying state is the maximally entangled state, up to some local isometries, through the observation of the maximal quantum violation of the CHSH inequality? Before solving this problem, we first apply a unitary on Bob's side and redefine the optimal observables:
\begin{equation}
\begin{aligned}
B_1^{\rm opt} = A_1^{\rm opt} =Z, \quad B_2^{\rm opt}=A_2^{\rm opt}=X.
\end{aligned}
\end{equation}
The form of the maximally entangled state under such a local unitary becomes:
\begin{equation}
|\psi^{\rm ref}\rangle^{\rm A'B'}:=\cos\frac{\pi}{8}|\Phi^-\rangle+\sin\frac{\pi}{8}|\Psi^+\rangle,
\end{equation}
where
 \begin{equation}
 \begin{aligned}
 & \left|\Phi^{-}\right\rangle=\frac{1}{\sqrt{2}}\left(|0\rangle_{\rm A'} \otimes|0\rangle_{\rm B'}-|1\rangle_{\rm A'} \otimes|1\rangle_{\rm B'}\right) \\
 & \left|\Psi^{+}\right\rangle=\frac{1}{\sqrt{2}}\left(|0\rangle_{\rm A'} \otimes|1\rangle_{\rm B'}+|1\rangle_{\rm A'} \otimes|0\rangle_{\rm B'}\right)
 \end{aligned}
 \end{equation}
are two of the Bell states.

The method of Ref.~\cite{Chen2021robustselftestingof} is inspired by the so-called \emph{swap method} proposed by Yang \emph{et al.}~\cite{Yang14}. In Ref.~\cite{Yang14}, to compute bounds on the fidelity, the authors choose a specific form of local isometries. Then, to obtain a device-independent result, they relax some characterized observables to unknown ones. In this work, with the method in Ref.~\cite{Chen2021robustselftestingof},  we firstly choose the identity maps $\mathsf{id}^{\rm A}: \mathcal{L}(\mathcal{H}^{\rm A})\rightarrow\mathcal{L}(\mathcal{H}^{\rm A'})$ and $\mathsf{id}^{\rm B}: \mathcal{L}(\mathcal{H}^{\rm B})\rightarrow\mathcal{L}(\mathcal{H}^{\rm B'})$ for both $\Lambda^{\rm A}$ and $\Lambda^{\rm B}$.  The choice has the same mathematical effect as the choice of the swap operator in Ref.~\cite{Yang14}. Namely,
\begin{equation}
\begin{aligned}
&\tr_{\rm AB}\Big\{U_{\rm \tiny{SWAP}}\Big[|0\rangle^{\rm A'}\langle 0|\otimes \rho^{\rm AB} \otimes |0\rangle^{\rm B'}\langle 0| \Big] U_{\rm \tiny{SWAP}}^\dag\Big\}\\
&=\mathsf{id}^{\rm A}\otimes\mathsf{id}^{\rm B}(\rho^{\rm AB}),
\end{aligned}
\end{equation}
where $U_{\rm \tiny{SWAP}}$ is the swap operator mapping the state $\rho^{\rm AB}$  from $\mathcal{L}(\mathcal{H}^{\rm AB})$ to $\mathcal{L}(\mathcal{H}^{\rm A'B'})$.

The Choi matrix $\Omega$ of identity operation $\mathsf{id}$ is
\begin{equation}
\Omega = \mathsf{id}\otimes \mathsf{id} (|\phi^+\rangle\langle \phi^+|)=|\phi^+\rangle\langle \phi^+|.
\end{equation}
Therefore, from Eq.~\eqref{Eq_duality}, we have
\begin{equation}
\begin{aligned}
&\mathsf{id}^{\rm A} \otimes \mathsf{id}^{\rm B} (\rho^{\rm AB}) =\\
& \tr_{\rm AB}\Big[ \Big( |\phi^+\rangle^{\rm AA'}\langle \phi^+|\otimes |\phi^+\rangle^{\rm BB'}\langle \phi^+|\Big)\Big( \rho^{\rm AB,\mathsf{T}}\otimes\mathbb{1}^{\rm A'B'} \Big)\Big].
\end{aligned}
\end{equation}
The subnormalized maximally entangled states $|\phi^+\rangle^{\rm AA'}\langle \phi^+|\in\mathcal{L}(\mathcal{H}^{\rm AA'})$ and $|\phi^+\rangle^{\rm BB'}\langle \phi^+|\in\mathcal{L}(\mathcal{H}^{\rm BB'})$ can be represented by the optimal observables $A_1^{\rm opt},A_2^{\rm opt}$ and $B_1^{\rm opt},B_2^{\rm opt}$, respectively:
\begin{equation}
\begin{aligned}
&|\phi^+\rangle^{\rm AA'}\langle \phi^+| = \\
&\frac{\id +A_1^{\rm opt}}{2}\otimes\left.|0\right\rangle\left\langle0|\right.+\frac{A_2^{\rm opt}-A_2^{\rm opt}A_1^{\rm opt}}{2}\otimes\left.|0\right\rangle\left\langle1|\right. \\
&+ \frac{A_2^{\rm opt}-A_1^{\rm opt}A_2^{\rm opt}}{2}\otimes\left.|1\right\rangle\left\langle0|\right.+\frac{\id -A_1^{\rm opt}}{2}\otimes\left.|1\right\rangle\left\langle1|\right.
\end{aligned}
\end{equation}
and
\begin{equation}
\begin{aligned}
&|\phi^+\rangle^{\rm BB'}\langle \phi^+| = \\
&\frac{\id +B_1^{\rm opt}}{2}\otimes\left.|0\right\rangle\left\langle0|\right.+\frac{B_2^{\rm opt}-B_2^{\rm opt}B_1^{\rm opt}}{2}\otimes\left.|0\right\rangle\left\langle1|\right. \\
&+\frac{B_2^{\rm opt}-B_1^{\rm opt}B_2^{\rm opt}}{2}\otimes\left.|1\right\rangle\left\langle0|\right.+\frac{\id -B_1^{\rm opt}}{2}\otimes\left.|1\right\rangle\left\langle1|\right.
\end{aligned}
\end{equation}
In a DI setting, the measurements are uncharacterized, therefore we relax the optimal observables to unknown Hermitian operators, i.e.,
\begin{equation}
A_x^{\rm opt}\rightarrow A_x,\quad B_y^{\rm opt}\rightarrow B_y.
\end{equation}
Consequently, a DI description of the fidelity between the reference state $|\psi^{\rm ref}\rangle$ and $\Lambda^{\rm A}\otimes\Lambda^{\rm B} (\rho^{\rm AB})$ is
\begin{equation}
\begin{aligned}
&F^{\rm DI} = \langle \psi^{\rm ref} | \Lambda^{\rm A} \otimes \Lambda^{\rm B} (\rho) |\psi^{\rm ref} \rangle\\
&=\langle \psi^{\rm ref} | \tr_{\rm AB}
[
(\id^{\rm A'} \otimes\rho^{\mathsf{T}}\otimes \id^{\rm B'})( \Omega^{\rm AA'}\otimes\Omega^{\rm BB'} )
] |\psi^{\rm ref} \rangle\\
&= \langle \psi^{\rm ref} | \tr_{\rm AB}
[
(\id^{\rm A'} \otimes\rho^{\mathsf{T}}\otimes \id^{\rm B'})( |\phi\rangle\langle\phi|\otimes|\phi\rangle\langle\phi| )
] |\psi^{\rm ref} \rangle\\
&=\sum_{i,j,k,l}{\ \left\langle\psi^{\rm ref}\right| \left (|i\right\rangle_{\rm A^\prime}\left\langle j|\right.\otimes\left.| k\right\rangle_{\rm B^\prime}\left\langle l |)\left|\psi^{\rm ref}\right\rangle\right.\ } \\
& \ \ \ \ \rm \cdot tr \mathit{ \left(\rho^{\mathsf{T}} \left.|i\right\rangle_{\rm A}\left\langle j|\right.\otimes\left.|k\right\rangle_{\rm B}\left\langle l|\right.\right)} \\
\end{aligned}
\label{Eq_FDI}
\end{equation}
The detailed derivation of the last equality can be found in Appendix \ref{SecApp_DI_F_MES}. Note that in the definition of robust self-testing,  Eq.~\eqref{Eq_Def_ST_state} holds for all states compatible with the observed correlation $\mathbf{P}_{\rm obs}$. Therefore, a lower bound $f$ can be operationally defined as the fidelity between the reference state and the \emph{worst} state $\rho^{\rm AB}$ (up to some local CPTPs). Namely, a choice for $f$ can be computed as:
\begin{equation}
\begin{aligned}
f := \min_{\rho^{\rm AB}}~F^{\rm DI} = \min_{\rho^{\rm AB}}~\langle \psi^{\rm ref} | \Lambda^{\rm A} \otimes \Lambda^{\rm B} (\rho) |\psi^{\rm ref} \rangle.
\end{aligned}
\label{Eq_min_FDI}
\end{equation}
On the other hand,  since the constraints of $F^{\rm DI}$ are more relaxed than those of $F$,  together with the minimization problem given in the above equation, we have
\begin{equation}
f:= \min_{\rho^{\rm AB}} F^{\rm DI} \leq F
\end{equation} 
Note that here we don't consider the optimization over all the CPTPs $\Lambda^{\rm A}\otimes \Lambda^{\rm B}$.

\begin{figure}[t]
\centering
\includegraphics[scale=0.5]{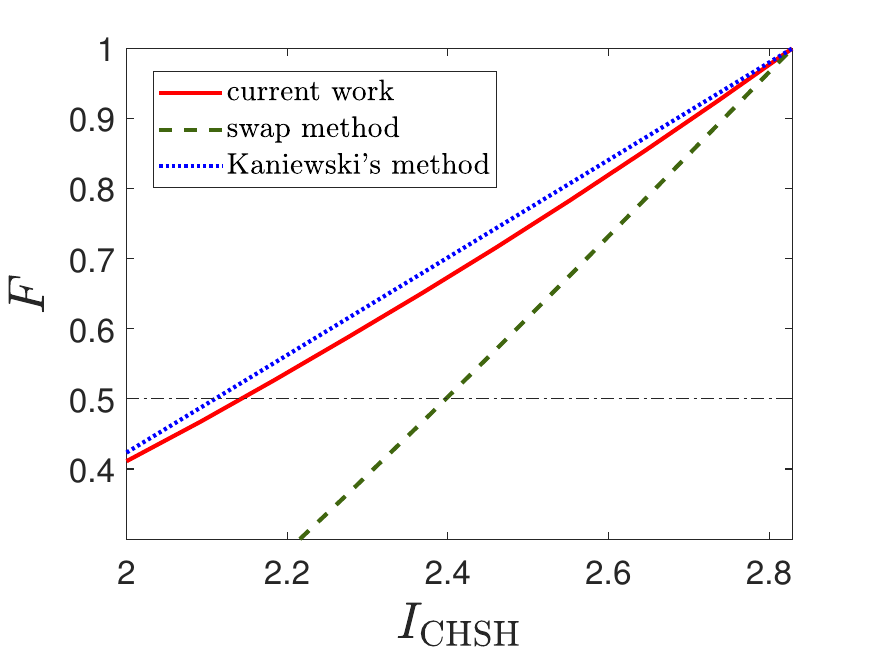}
\caption{
Robust self-testing of maximally entangled states. Given a CHSH inequality violation, we use Eq.~\eqref{Eq_min_F_SDP} to compute a lower bound on the fidelity. To carry out the computation, we use the $3$rd level of the hierarchy, defined in Eq.~\eqref{EqApp_Sequence_CHSH}. To compare our method with other works, we also attach the result computed by the swap method~\cite{Yang14} and the Kaniewski's method of Ref.~\cite{Kaniewski16}.
}
\label{Fig_CHSH}
\end{figure}

Now, we focus on solving the minimization problem of Eq.~\eqref{Eq_min_FDI}. The problem can be formulated as
\begin{equation}
\begin{aligned}
\min_{\rho^{\rm AB}}~~&\langle \psi^{\rm ref} | \Lambda^{\rm A} \otimes \Lambda^{\rm B} (\rho^{\rm AB}) |\psi^{\rm ref} \rangle\\
\text{such that}~~& \rho^{\rm AB}~~\text{is compatible with~~}\mathbf{P}_{\rm obs}.
\end{aligned}
\end{equation}
Or, equivalently, 
\begin{equation}
\begin{aligned}
\min_{\mathbf{P}}~~&\langle \psi^{\rm ref} | \Lambda^{\rm A} \otimes \Lambda^{\rm B} (\rho^{\rm AB}) |\psi^{\rm ref} \rangle\\
\text{such that}~~
& \mathbf{P} = \mathbf{P}_{\rm obs},\\
& \mathbf{P}\in\mathcal{Q},
\end{aligned}
\end{equation}
where $\mathcal{Q}$ is the set of quantum correlations. The last constraint is hard to characterize, but we can use the technique of semidefinite programming relaxation for quantum correlations~\cite{NPA2008,Doherty08,Tavakoli23} to characterize a \emph{superset}:
\begin{equation}
\begin{aligned}
\min~~&\langle \psi^{\rm ref} | \Lambda^{\rm A} \otimes \Lambda^{\rm B} (\rho) |\psi^{\rm ref} \rangle\\
\text{such that}~~
& \mathbf{P} = \mathbf{P}_{\rm obs},\\
& \Gamma\succeq 0,
\end{aligned}
\label{Eq_min_F_SDP}
\end{equation}
where
\begin{equation}
\Gamma = \Gamma(\rho,\mathcal{S}) := \sum_{ij}|i\rangle\langle j|\tr\Big( S_j^\dag S_i \rho  \Big)
\label{Eq_moment_matrix}
\end{equation}
is the moment matrix of the state $\rho$ associated with the sequence $\mathcal{S}$, which is composed of POVMs $E_{a|x}, E_{b|y}$ and their products. The minimization is taken over all unknown variables in $\Gamma$.

In Eq.~\eqref{Eq_min_F_SDP}, the objective function is a polynomial of expectation values of POVMs and their products, such as $\tr(\rho E_{a|x}\otimes E_{b|y}), \tr(\rho E_{a|x}E_{a'|x'}\otimes E_{b|y}), \tr(\rho E_{a|x}\otimes E_{b|y}E_{b'|y'})$ etc. (see Eq.~\eqref{Eq_FDI}). The moment matrix $\Gamma$ contains these terms. Therefore, Eq.~\eqref{Eq_min_F_SDP} is a semidefinite program (SDP)~\cite{BoydBook}, which can be solved numerically by some computer packages~\cite{Lofberg2004,SDPT3}. We implement the computation of the SDP and attach the result in Fig.~\ref{Fig_CHSH}.

\section{Self-testing of the two-qubit partially entangled states}\label{Sec_ST_PES}
In this section, we show that the method can be applied to self-testing of a family of two-qubit partially entangled states:
\begin{equation}
|\psi^{\rm ref}\rangle = \cos\theta|00\rangle + \sin\theta |11\rangle.
\end{equation}
The first few steps are the same as the previous section. That is, first, we identify a Bell inequality which can be maximally violated by $|\psi^{\rm ref}\rangle$. An instance is the so-called \emph{tilted CHSH inequality}~\cite{Acin12,Yang13,Bamps15}:
\begin{equation}
\begin{aligned}
&I_{\rm tilted-CHSH}:= \alpha\langle A_1\rangle + \\
&\langle A_1 B_1\rangle + \langle A_1 B_2\rangle + \langle A_2 B_1\rangle - \langle A_2 B_2\rangle\leq 2 + \alpha.
\end{aligned}
\end{equation}
where $0\leq \alpha<2$. The maximal quantum violation $\sqrt{8+2\alpha^2}$ can be achieved by the above quantum state and the following measurements:
\begin{equation}
\begin{aligned}
A_1^{\rm opt} =Z, \quad B_1^{\rm opt} = \cos\mu Z + \sin\mu X,\\
A_2^{\rm opt} =X, \quad B_2^{\rm opt} = \cos\mu Z - \sin\mu X,
\end{aligned}
\end{equation}
where the relations between $\theta$, $\alpha$, and $\mu$ are
\begin{equation}
\tan\mu = \sin 2\theta=\sqrt{\frac{4-\alpha^2}{4+\alpha^2}}.
\end{equation}
In the previous section, we apply a unitary on Bob's side to make the optimal observables the same as Alice's. Here, instead, we introduce additional two observables $B_3^{\rm opt}$ and $B_4^{\rm opt}$ such that~\cite{Yang14,Bancal15}
\begin{equation}
\begin{aligned}
B_3^{\rm opt}\left( \frac{B_1^{\rm opt} + B_2^{\rm opt}}{\cos\mu} \right)\succeq 0,\quad
B_4^{\rm opt}\left( \frac{B_1^{\rm opt} - B_2^{\rm opt}}{\sin\mu} \right)\succeq 0.
\end{aligned}
\label{Eq_PSD_Bob}
\end{equation}
The idea behind is that given an operator $B$, there is a unitary operator $U$ such that $UB$ is positive semidefinite. Besides, if $B$ is unitary, then $U=B^\dag$. As a result, since $(B_1^{\rm opt} + B_2^{\rm opt})/\cos\mu$ and $(B_1^{\rm opt} - B_2^{\rm opt})/\sin\mu$ are unitary, we have $B_3^{\rm opt}=Z$ and $B_4^{\rm opt}=X$. This allows us to keep $B_3^{\rm opt}$ and $B_4^{\rm opt}$ unitary when they are relaxed to known observables.

\begin{figure*}[htbp]
\begin{minipage}[b]{1\linewidth}
    \centering
    \subfigure[]{\includegraphics[scale=0.5]{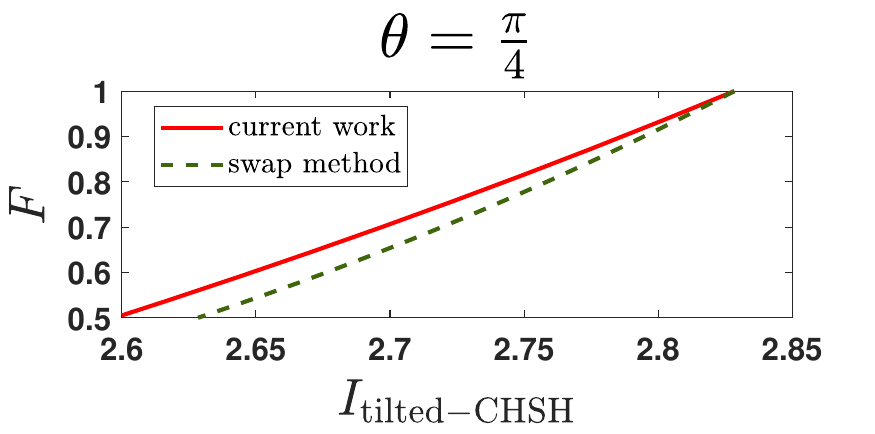}}
    \hfill
    \subfigure[]{\includegraphics[scale=0.5]{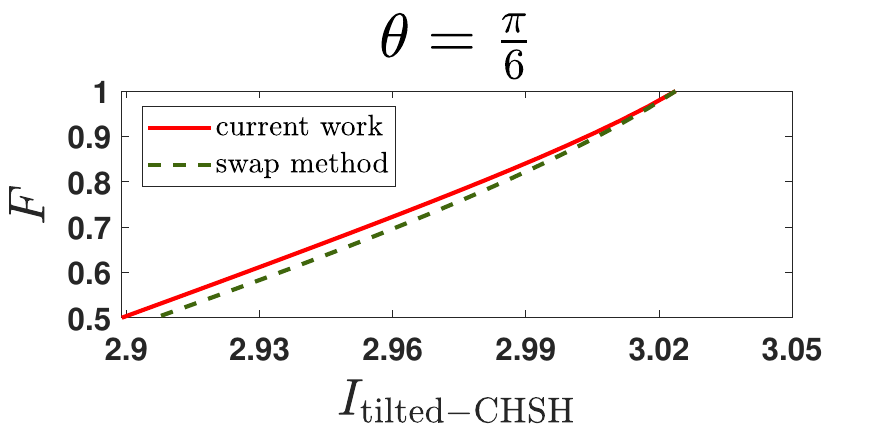}}
    \hfill
    \subfigure[]{\includegraphics[scale=0.5]{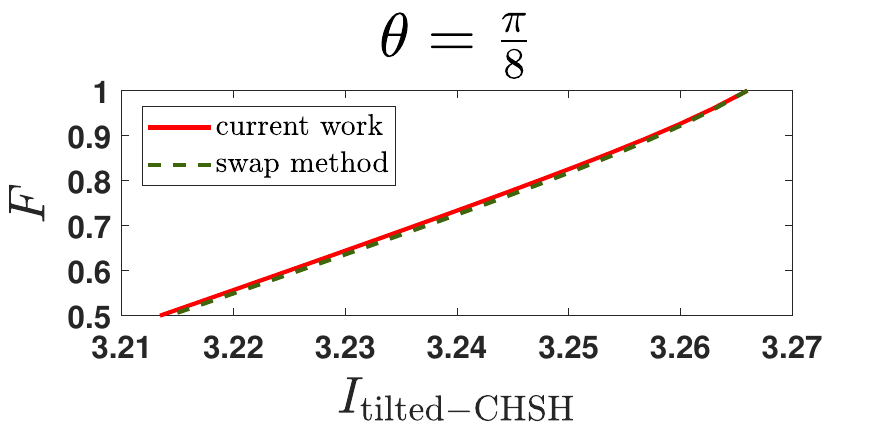}}
    \caption{The figure shows the bounds of self-testing family of pure entangled quantum states, parameterized by different values of $\theta$. The axis $I_{\rm tilted- CHSH}$ in each figure is the violation of the tilted CHSH inequality.}
    \label{Fig_minF_tCHSH}
\end{minipage}
\end{figure*}

Now, we again choose the identity operations for Alice's and Bob's CPTP maps. The associated Choi matrix is the subnormalized maximally entangled state and can be represented as follows:
\begin{equation}
\begin{aligned}
& \Omega^{\rm AA'} = \\
&\frac{\id +A_1^{\rm opt}}{2}\otimes\left.|0\right\rangle\left\langle0|\right.+\frac{A_2^{\rm opt}-A_2^{\rm opt}A_1^{\rm opt}}{2}\otimes\left.|0\right\rangle\left\langle1|\right. \\
&+\frac{A_2^{\rm opt}-A_1^{\rm opt}A_2^{\rm opt}}{2}\otimes\left.|1\right\rangle\left\langle0|\right.+\frac{\id -A_1^{\rm opt}}{2}\otimes\left.|1\right\rangle\left\langle1|\right.
\end{aligned}
\end{equation}
and
\begin{equation}
\begin{aligned}
& \Omega^{\rm BB'} = \\
&\frac{\id +B_3^{\rm opt}}{2}\otimes\left.|0\right\rangle\left\langle0|\right.+\frac{B_4^{\rm opt}-B_4^{\rm opt}B_3^{\rm opt}}{2}\otimes\left.|0\right\rangle\left\langle1|\right. \\
&+\frac{B_4^{\rm opt}-B_3^{\rm opt}B_4^{\rm opt}}{2}\otimes\left.|1\right\rangle\left\langle0|\right.+\frac{\id -B_3^{\rm opt}}{2}\otimes\left.|1\right\rangle\left\langle1|\right.
\end{aligned}
\end{equation}
In a DI setting, the measurements are not characterized, therefore we relax the optimal observables $A_x^{\rm opt},B_y^{\rm opt}$ to unknown Hermitian operators $A_x,B_y$. Again, with the procedure of Eq.~\eqref{Eq_FDI}, a DI description of the fidelity can be obtained. Indeed, the form will be exactly the same as Eq.~\eqref{EqApp_FDI_Derivation}, with substituting $B_1$ and $B_2$ with $B_3$ and $B_4$, respectively.
Consequently, given a quantum violation of $I_{\rm tilted-CHSH}$, a lower bound $f$ on the fidelity can be computed by the following SDP:
\begin{equation}
\begin{aligned}
\min~~&F^{\rm DI}\\
\text{such that}~~&\mathbf{P} = \mathbf{P}_{\rm obs},\\
&\Gamma\succeq 0,\\
&\Gamma_{\rm L}\Big[ \frac{B_3(B_1+B_2)}{\cos\mu},\mathcal{S}' \Big]\succeq 0,\\
&\Gamma_{\rm L}\Big[ \frac{B_3(B_1-B_2)}{\sin\mu},\mathcal{S}' \Big]\succeq 0.
\end{aligned}
\label{Eq_minF_pure_ent_state}
\end{equation}
Here, $\Gamma_{\rm L}$ is the \emph{localizing matrix}~\cite{Pironio10,Yang14,Bancal15}, defined as
\begin{equation}
\Gamma_{\rm L}[B,\mathcal{S}']:=\sum_{ij}\text{tr}[(S'_j)^\dag B S'_i \rho],
\end{equation}
and it is positive semidefinite if $B$ is positive semidefinite. Therefore, the positive semidefiniteness of the last two constraints in the above SDP is relaxed from the constraints given by Eq.~\eqref{Eq_PSD_Bob}. The sequence $\mathcal{S}'$ here is not necessarily the same as the sequence $\mathcal{S}$ used for constructing the standard moment matrix $\Gamma$. Instead, we require that 1) $\Gamma_{\rm L}$ generated by $\mathcal{S}'$ contains all the moment terms of the fidelity $F^{\rm DI}$ and 2) $\mathcal{S}'\subseteq\mathcal{S}$. In Fig.~\ref{Fig_minF_tCHSH}, we present the result of robust self-testing of different pure entangled states, parameterized by values of $\theta$.

\section{Summary and Discussion}\label{Sec_summary}
In this work, we consider the previous method of self-testing steerable quantum assemblages and generalize this method to self-test entangled quantum states. In each Bell scenario under consideration, we obtain the perfect self-testing result for the maximal quantum violation of the Bell inequality. We also obtain non-trivial fidelity even when the Bell inequality violation is not maximal, which means that the self-testing result is robust.

We list some open questions as follows. First, can this method be applied to the self-testing of complex-valued quantum states? Second, are there other forms of the Choi matrix that can be used to increase the bounds of the fidelity? Finally,  we expect that the method can be applied to other self-testing tasks, such as self-testing of channels and measurements.

\acknowledgements S.-L.~C.~acknowledges the support of the National Science and Technology Council (NSTC) Taiwan (Grant No. NSTC 111-2112-M-005-007-MY4) and National Center for Theoretical Sciences Taiwan (Grant No.~NSTC 113-2124-M-002-003)


\bibliography{bib_self_testing_states}

\begin{thebibliography}{42}
\bibitem [{FN1()}]{FN1}%
  \BibitemOpen
  \href@noop {} {}\bibinfo {note} {We add a subscript ``obs'' on $\mathbf{P}$
  when emphasizing that the entries of the correlation $\mathbf{P}$ are given
  known data, not unknown variables. Otherwise there is no physical difference
  between $\mathbf{P}$ and $\mathbf{P}_{\rm obs}$.}\BibitemShut {Stop}%
\end{thebibliography}

\clearpage
\onecolumngrid
\appendix

\section{Derivation of the DI fidelity of the reference states}\label{SecApp_DI_F_MES}

In Eq.~\eqref{Eq_FDI}, the detailed representation of the DI fidelity of the two-qubit maximally entangled state is as follows:
\begin{equation}
\begin{aligned}
&F^{\rm DI}_{\rm chsh} = \frac{1}{8} \ [ \  \cos^2 \frac{\pi}{8} \left(\left\langle A_1\right\rangle +\left\langle B_1\right\rangle +\left\langle A_1B_1\right\rangle +\mathbb{1}\right) + \cos\frac{\pi}{8}\sin\frac{\pi}{8} \left(\left\langle B_2\right\rangle +\left\langle A_1B_2\right\rangle -\left\langle B_2B_1\right\rangle -\left\langle A_1B_2B_1\right\rangle \right) \\
&+ \cos\frac{\pi}{8}\sin\frac{\pi}{8} \left(\left\langle B_2\right\rangle +\left\langle A_1B_2\right\rangle -\left\langle B_1B_2\right\rangle -\left\langle A_1B_1B_2\right\rangle \right) + \sin^2 \frac{\pi}{8} \left(\left\langle A_1\right\rangle -\left\langle B_1\right\rangle -\left\langle A_1B_1\right\rangle +\mathbb{1}\right) \\
&+ \cos\frac{\pi}{8}\sin\frac{\pi}{8} \left(\left\langle A_2\right\rangle -\left\langle A_2A_1\right\rangle +\left\langle A_2B_1\right\rangle -\left\langle A_2A_1B_1\right\rangle \right) 
- \cos^2 \frac{\pi}{8} \left(\left\langle A_2B_2\right\rangle -\left\langle A_2A_1B_2\right\rangle -\left\langle A_2B_2B_1\right\rangle +\left\langle A_2A_1B_2B_1\right\rangle \right) \\
&+ \sin^2 \frac{\pi}{8} \left(\left\langle A_2B_2\right\rangle -\left\langle A_2A_1B_2\right\rangle -\left\langle A_2B_1B_2\right\rangle +\left\langle A_2A_1B_1B_2\right\rangle \right) 
- \cos\frac{\pi}{8}\sin\frac{\pi}{8} \left(\left\langle A_2\right\rangle -\left\langle A_2A_1\right\rangle -\left\langle A_2B_1\right\rangle +\left\langle A_2A_1B_1\right\rangle \right) \\
&+ \cos\frac{\pi}{8}\sin\frac{\pi}{8} \left(\left\langle A_2\right\rangle -\left\langle A_1A_2\right\rangle +\left\langle A_2B_1\right\rangle -\left\langle A_1A_2B_1\right\rangle \right) 
+ \sin^2 \frac{\pi}{8} \left(\left\langle A_2B_2\right\rangle -\left\langle A_1A_2B_2\right\rangle -\left\langle A_2B_2B_1\right\rangle +\left\langle A_1A_2B_2B_1\right\rangle \right) \\
&- \cos^2 \frac{\pi}{8} \left(\left\langle A_2B_2\right\rangle -\left\langle A_1A_2B_2\right\rangle -\left\langle A_2B_1B_2\right\rangle +\left\langle A_1A_2B_1B_2\right\rangle \right) 
- \cos\frac{\pi}{8}\sin\frac{\pi}{8} \left(\left\langle A_2\right\rangle -\left\langle A_1A_2\right\rangle -\left\langle A_2B_1\right\rangle +\left\langle A_1A_2B_1\right\rangle \right) \\
&+ \sin^2 \frac{\pi}{8} \left(\left\langle B_1\right\rangle -\left\langle A_1\right\rangle -\left\langle  A_1B_1\right\rangle +\mathbb{1}\right) 
- \cos\frac{\pi}{8}\sin\frac{\pi}{8} \left(\left\langle B_2\right\rangle -\left\langle  A_1B_2\right\rangle -\left\langle B_2B_1\right\rangle +\left\langle A_1B_2B_1\right\rangle \right) \\
&- \cos\frac{\pi}{8}\sin\frac{\pi}{8} \left(\left\langle B_2\right\rangle -\left\langle A_1B_2\right\rangle -\left\langle B_1B_2\right\rangle +\left\langle A_1B_1B_2\right\rangle \right) 
+ \cos^2 \frac{\pi}{8} \left(\left\langle A_1B_1\right\rangle -\left\langle B_1\right\rangle -\left\langle A_1\right\rangle +\mathbb{1}\right) \ ]
\end{aligned}
\label{EqApp_FDI_Derivation}
\end{equation}
We implement the computation of the SDP~\eqref{Eq_min_F_SDP} by plugging the above into the objective function and define the $3$rd level of sequence of operators, namely,
\begin{equation}
\begin{aligned}
\mathcal{S}:=\left\{S_i\right\}= \openone \cup \mathcal{S}^{(1)}  \cup \mathcal{S}^{(2)}  \cup \mathcal{S}^{(3)},
\end{aligned}
\label{EqApp_Sequence_CHSH}
\end{equation}
where $ \mathcal{S}^{(\ell)}$ is the set of $\ell$th-order products of observables $A_x$ and $B_y$.

For the family of pure entangled states, the fidelity is expressed as follows:
\begin{equation}
\begin{aligned}
&F^{\rm DI}_{\rm tilted}  = \frac{1}{4} \ [ \cos^2 \theta (\mathbb{1} +  \left\langle A_1 \right\rangle + \left\langle B_3 \right\rangle + \left\langle A_1 B_3 \right\rangle) + \cos\theta\sin\theta ( \left\langle A_2 B_4 \right\rangle - \left\langle A_2 B_4 B_3 \right\rangle - \left\langle A_2 A_1 B_4 \right\rangle + \left\langle A_2 A_1 B_4 B_3 \right\rangle) \\
&+ \cos\theta\sin\theta ( \left\langle A_2 B_4 \right\rangle - \left\langle A_2 B_3 B_4 \right\rangle - \left\langle A_1 A_2 B_4 \right\rangle + \left\langle A_1 A_2 B_3 B_4 \right\rangle) + \sin^2 \theta (\mathbb{1} - \left\langle A_1 \right\rangle - \left\langle B_3 \right\rangle + \left\langle A_1 B_3 \right\rangle) \ ]
\end{aligned}
\label{EqApp_F_tilted_DI_Derivation}
\end{equation}
The computation is carried out by running the SDP~\eqref{Eq_minF_pure_ent_state} with the following sequence:
\begin{equation}
\begin{aligned}
\left\{S'_i\right\}=\mathcal{S}_{\rm CHSH}\cup \{ 
&B_3B_4,B_4B_3,  B_1B_4,  B_4B_1,B_3B_1,B_1B_3, 
A_1B_3, A_2B_3, A_1B_4,A_2B_4, \\
&B_3B_4B_3,B_4B_3B_4, A_1B_3B_4,
A_2B_3B_4,A_1B_4B_3,A_2B_4B_3\},
\end{aligned}
\end{equation}
where $\mathcal{S}_{\rm CHSH}$ is the sequence of operators we just used in self-testing of maximally entangled states, i.e., the form described in Eq.~\eqref{EqApp_Sequence_CHSH}.

\end{document}